\documentclass{mem}
\usepackage{natbib}\usepackage{txfonts}\usepackage{balance}
\usepackage{graphicx}
\usepackage[a4paper]{hyperref}
\idline{79}{0}

\begin{document}

\title{The cosmological co-evolution of supermassive black holes, AGN and galaxies}

\author{
Federico Marulli\inst{1}, 
Silvia Bonoli\inst{2},
Enzo Branchini\inst{3},
Lauro Moscardini\inst{1,4} and \\
Volker Springel\inst{2}
}
\offprints{F. Marulli}

\institute{
Dipartimento di Astronomia, Universit\`a degli Studi di Bologna, 
via Ranzani 1, I-40127 Bologna, Italy \email{federico.marulli3@unibo.it}
\and
Max-Planck-Institut f\"ur Astrophysik, Karl-Schwarzschild Strasse 1,
D-85740 Garching, Germany
\and
Dipartimento di Fisica, Universit\`a 
degli Studi ``Roma Tre'', via della Vasca Navale 84, I-00146 Roma,
Italy
\and
INFN, Sezione di Bologna, viale Berti Pichat 6/2, I-40127 Bologna,
Italy
}

\authorrunning{Marulli et al.}
\titlerunning{Modelling BH properties and the AGN luminosity function}

\abstract{We model the cosmological co-evolution of galaxies and their
  central supermassive black holes (BHs) within a semi-analytical
  framework developed on the outputs of the Millennium Simulation
  \citep{croton2006,delucia2007}. In this work, we analyze the model
  BH scaling relations, fundamental plane and mass function, and
  compare them with the most recent observational data. Furthermore, we
  extend the original code developed by \citet{croton2006} to follow
  the evolution of the BH mass accretion and its conversion into
  radiation, and compare the derived AGN bolometric luminosity
  function with the observed one. We find, for the most part, a very
  good agreement between predicted and observed BH properties.
  Moreover, the model is in good agreement with the observed AGN
  number density in $0\leq z\leq5$, provided it is assumed that the cold gas
  fraction accreted by BHs at high redshifts is larger than at low
  redshifts \citep{marulli2008}.  \keywords{ AGN: general -- galaxies:
    formation -- galaxies: active -- cosmology: theory -- cosmology:
    observations} }

\maketitle{}

\section {The model} \label{sec:description}
Our cosmological model for the co-evolution of DM haloes, galaxies and
their central BHs consists of three ingredients: i) a numerical N-body
simulation, the Millennium Run, to obtain the merger history of the DM
haloes and subhaloes within the framework of the $\Lambda$CDM model,
ii) a set of analytic prescriptions to trace the formation and
evolution of galaxies and iii) a set of recipes to follow the BH
accretion history and the AGN phenomenon.  This model is described in
detail in \citet{croton2006}, \citet{delucia2007} and \citet
{marulli2008}. In the following, we just give a brief description of
the assumptions introduced to describe the BH and AGN evolution.

\subsection{Black Holes and Active Galactic Nuclei}

\subsubsection{Quasar Mode}
As it is well established that galaxy major mergers cannot constitute
the only trigger to accretion episodes in the local BH population
\citep[see e.g.][and reference therein]{marulli2007}, we assume that
BHs can accrete mass after every galaxy mergers, both through
coalescence with another BH and by accreting cold gas, the latter
being the dominant accretion mechanism \citep[see e.g.][and reference
  therein]{marulli2006}. For simplicity, the BH coalescence is
modelled as a direct sum of the progenitor masses, thus ignoring
gravitational wave losses. We assume that the gas mass accreted during
a merger is proportional to the total cold gas mass present, but with
an efficiency which is lower for smaller mass systems and for unequal
mergers.  This kind of accretion is also closely associated with
starbursts, which occur concurrently. We do not model feedback from
the quasar activity, but it can be approximately represented by an
enhanced effective feedback efficiency for the supernovae associated
with the intense starburst.

The evolution of an active BH is described as a two-stage process of a
rapid, Eddington-limited growth up to a peak BH mass, preceded and
followed by a much longer quiescent phase with lower Eddington ratios.
In this latter phase, the average time spent by AGN per logarithmic
luminosity interval can be approximated as in \citet{hopkins2005}.

As discussed in details in \citet{marulli2008}, with the original
semi-analytic recipes introduced by \citet{croton2006} to follow the
BH mass accretion, the model underestimates the number density of
luminous AGN at high redshifts, independently of the assumptions
introduced to describe the accretion efficiency, the Eddington factor,
or the BH seed masses. Significant improvement can be obtained by
simply assuming an accretion efficiency that increases with the
redshift.  In the following Section, we will focus on the predictions
of this new model, which we called `best model' in
\citet{marulli2008}.

\subsubsection{Radio Mode}
We assume that, when a static hot halo has formed around a galaxy, a
fraction of the hot gas continuously accretes onto the central BH,
causing a low-energy `radio' activity in the galaxy centre.  This
accretion rate is typically orders-of-magnitude below the Eddington
limit, so that the total mass growth of BHs in the radio relative to
the quasar mode is negligible. It is also assumed that the radio mode
feedback injects energy efficiently into the surrounding medium, which
can reduce or even stop the cooling flow in the halo centres.  In this
scenario, the effectiveness of radio AGN in suppressing cooling flows
is greatest at late times and for large values of the BH mass, which
is required to successfully reproduce the luminosities, colours and
clustering of low-redshift bright galaxies.

\section {Models vs. Observations} \label{sec:comparison}
In this Section, we show the most interesting results about the
comparison between our model predictions and several observed
statistical properties of BHs and AGN.  In all the Figures, black and
grey symbols show the observational data, while red and blue ones show
the model predictions.

\begin{figure}
  \includegraphics[width=0.45\textwidth]{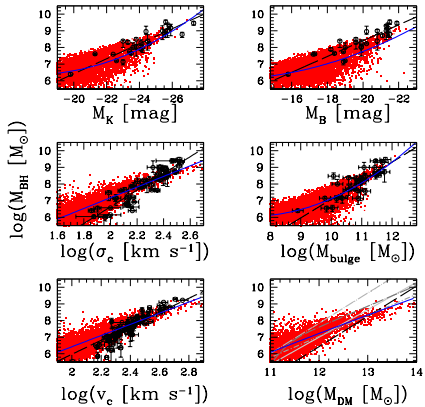} 
\caption{ Scaling relations between the masses of the central BHs in
  the simulated galaxies with six different properties of their hosts:
  the K- and B-band bulge magnitude (top left and right panels,
  respectively), the bulge velocity dispersion and mass (central left
  and right panels, respectively), the circular velocity of the galaxy
  (bottom left panel) and the virial mass of the DM halo (bottom right
  panel). Red and black dots represent model predictions and
  observations, respectively. Solid blue and dashed black and grey
  lines show the best fit to the model predictions and to the
  observational datasets, respectively.}
  \label{fig:scale1}
\end{figure}

Figure \ref{fig:scale1} shows the correlation between the masses of
the model BHs with six properties of their hosts: the K- and B-band
bulge magnitude (${\rm M}_{\rm B}$ and ${\rm M}_{\rm K}$)
\citep{marconi2003}, the bulge mass and velocity dispersion (${\rm
  M}_{\rm bulge}$ and $\sigma_{\rm c}$)
\citep{haring2004,ferrarese2005}, the circular velocity of the galaxy
and the virial mass of the DM halo (${\rm V}_{\rm c}$ and ${\rm
  M}_{\rm DM}$) \citep{ferrarese2002,baes2003,shankar2006}.

\begin{figure}
  \includegraphics[width=0.45\textwidth]{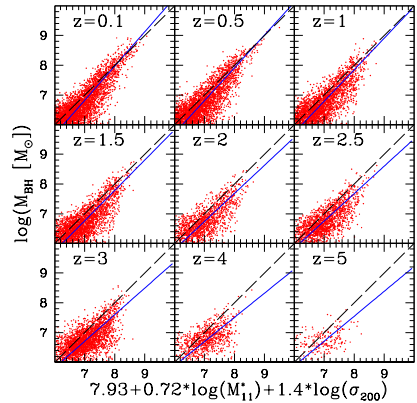}
  \caption{The BH fundamental plane in the redshift range $0.1\leq
    z\leq5$. The red dots are the model outputs, while the blue solid
    lines show the best fits to them. The black dashed lines show the
    predictions of \citet{hopkins2007a}. The galaxy stellar mass,
    $M_{11}^*$, is given in units of $10^{11}M_\odot$, while the bulge
    velocity dispersion, $\sigma_{200}$, is in units of 200 km
    $s^{-1}$.}
  \label{fig:bhfp}
\end{figure}

In Figure \ref{fig:bhfp}, we compare the BH fundamental plane relation
in the redshift range $0.1\leq z\leq5$ predicted by our model with
that obtained by \citet{hopkins2007a}, using both observational data
and the outputs of hydrodynamical simulations of galaxy merger. 

\begin{figure}
  \includegraphics[width=0.45\textwidth]{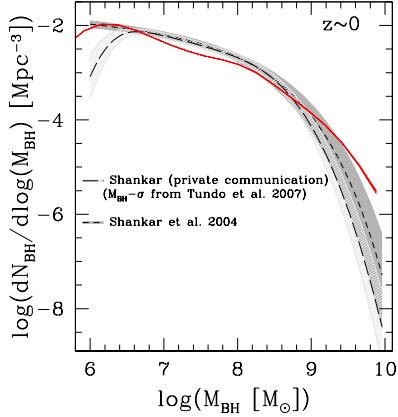}
  \caption{The model BH mass function (red line) compared with the
    one observationally derived by \citet{shankar2004} (dark grey
    area), and with the new one obtained by Shankar (private
    communication) (light grey area) using the $M_{\rm BH}-\sigma$
    relation by \citet{tundo2007}.}
  \label{fig:bhmf}
\end{figure}

In Figure \ref{fig:bhmf}, we show the BH mass function predicted by
our model with those observed by \citet{shankar2004} and by Shankar
(private communication) at $z\sim 0$.  In neither case the BH masses
were determined directly: \citet{shankar2004} derive the BH mass from
the observed ${\rm M}_{\rm BH}-L_{\rm bulge}$ relation, while Shankar
(private communication) uses the ${\rm M}_{\rm BH}-\sigma_{\rm c}$
relation of \citet{tundo2007}.

\begin{figure}
\includegraphics[width=0.45\textwidth]{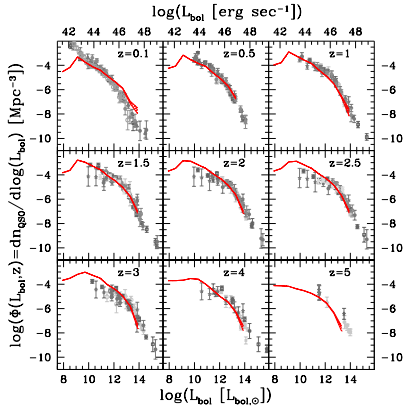}
\caption{The model AGN bolometric luminosity functions (red lines)
  compared with several observed ones (grey dots). The bolometric
  corrections adopted and the datasets considered are the ones
  discussed in \citet{hopkins2006}}
\label{fig:lf}
\end{figure}

Finally, in Figure \ref{fig:lf} we compare the AGN bolometric
luminosity function predicted by our model with several observed ones
in different bands.  The bolometric corrections adopted and the
datasets considered are the ones discussed in \citet{hopkins2006}.

\section {Conclusions} \label{sec:conclusions}

\noindent The main results  of our analysis are as follows:

{\em (i)} The semi-analytic model is able to reproduce the observed BH
scaling relations over the whole range of BH masses and galaxy
properties probed by observations.  The intrinsic scatter in the model
is significantly larger than in the data, a mismatch that can be
accounted for by adopting the observational selection criteria to
obtain a mock BH catalogue with similar characteristics as the
observed one.

{\em (ii)} We find evidence that a quadratic relationship provides a
significantly better fit to some of the model scaling relationships than
a linear one, as already noticed by \citet{wyithe2006}.

{\em (iii)} Our model also matches the BH fundamental plane relation
derived by \citet{hopkins2007a}, and predicts very little evolution of
this plane, at least out to $z\sim 3$.

{\em (iv)} The model BH mass function is in good agreement with the
observed one within the mass range accessible by observations, 
except on the range $\sim10^7-10^9 M_\odot$, in which the number density
predicted by the model is smaller than the observed one.

{\em (v)} Model predictions for the BH mass function, scaling relations
and fundamental plane relation are basically unaffected when using
different prescriptions for the AGN lightcurves of individual quasar
events. 

{\em (vi)} The model underestimates the number density of luminous AGN
at high redshifts, independently of the lightcurve model adopted. We
were not able to eliminate this mismatch by simply modifying the
accretion efficiency, the Eddington factor or the BH seed mass (when
considered in physically plausible ranges).  A simple, {\em ad hoc}
increase of the mass fraction accreted during the {\em quasar mode} at
high redshifts can indeed remedy the problem. However, this solution is
not unique as several high-redshift modifications to the original
model, like new mechanisms that trigger BH activity in addition to
galaxy merging or more efficient gas cooling resulting in a larger
reservoir of cold gas, can be advocated to bring the predictions in
line with observations.  However, it remains to be seen whether any of
these alternatives is physically plausible.

\bibliographystyle{aa}
\bibliography{bib}

\label{lastpage}

\end{document}